# High pressure study of transport properties in $Co_{1/3}NbS_2$

N. Barišić[1,2,*], I. Smiljanić[1], P. Popčević[1], A. Bilušić[1,3], E. Tutiš[1], A. Smontara[1], H. Berger[4], J. Jaćimović[4], O. Yuli[4] and L. Forró[4]

[1]*Institute of Physics, Bijenička c. 46, HR–10000 Zagreb, Croatia*

[2]*1. Physikalisches Institut, Universität Stuttgart, D-70550 Stuttgart, Germany*

[3]*Faculty of Mathematics, Natural Sciences and Education, University of Split, Nikole Tesle 12, HR–21000 Split, Croatia*

[4]*École Polytechnique Fédérale de Lausanne, IPMC/SB, EPFL, CH-1015 Lausanne, Switzerland*

*Corresponding author: barisic@ifs.hr

**Abstract**

This is the first study of the effect of pressure on transition metal dichalcogenides intercalated by atoms that order magnetically. $Co_{1/3}NbS_2$ is a layered system where the intercalated Co atoms order antiferromagnetically at $T_N = 26$ K at ambient pressure. We have conducted a detailed study of dc-resistivity ($\rho$), thermoelectric power (*S*) and thermal conductivity ($\kappa$). We found that at ambient pressure the magnetic transition corresponds to a well pronounced peak in $dS/dT$, as well as to a kink in the dc-resistivity. The effect of ordering on the thermal conductivity is rather small but, surprisingly, more pronounced in the lattice contribution than in the electronic contribution to $\kappa$. Under pressure, the resistivity increases in the high temperature range, contrary to all previous measurements in other layered transition metal dichalcogenides (TMD). In the low temperature range, the strong dependences of thermopower and resistivity on pressure are observed below $T_N$, which, in turn, also depends on pressure at rate of $dT_N/dp \sim -1$ K/kbar. Several possible microscopic explanations of the reduction of the ordering temperature and the evolution of the transport properties with pressure are discussed.

1. INTRODUCTION

The interest in layered electronic materials is fueled by their capability to host peculiar electronic phases. Among them, transition metal dichalcogenides (TMDs) (of general formula $TX_2$, where T – transition element and X – S, Se, Te) are well known as the material where charge density waves, superconductivity, metallic and excitonic phases compete for the ground state.[1,2,3] The magnetic ordering, related to ions intercalated in between TMD layers coupled by the van der Waals forces, has also attracted great interest in the past. $M_{1/3}TX_2$, (M=Co, Fe, Ni, Mn, …) compounds were of a particular interest because the magnetic ions in their layers form a triangular lattice, which is known to be prone to frustration . Experimentally, several types of magnetic orderings was found in these materials, thought to be driven by competing interactions.[4]

The sensitivity of the electronic properties of TMD's to pressure in general is well known. Several recent publications have shown that pressure is an important factor, affecting the electronic phases of TMD's, leading to a rich phase diagram.[5,6] Until now pressure was not used to disclose the physical properties of TMD's intercalated by ions. Here we study the effect of pressure on transport properties and the ordering of the system.

The material under consideration, $Co_{1/3}NbS_2$, is known to order magnetically at 26 K at ambient pressure.[4,7,8] The measurements at ambient pressure, the temperature dependence and the manifestation of the magnetic ordering on a set of transport properties are presented first. This is followed by a study of the material under pressure, which shows the peculiar effect of pressure on transport and ordering at low temperature, as well as its unexpected influence on the resistivity in the high temperature range.

The paper is as follows: The next section introduces the material and its parent compound. Section 3 gives the experimental details. The experimental results are presented and partly discussed in Section 4, first at ambient, and then under elevated pressure. The possible mechanisms behind the observed pressure dependence are discussed in Section 5.

2. MATERIAL

TMD's structure is sandwich-like with tight intra-layer couplings and loose van der Walls inter-layer coupling. Consequently, they are electronically highly anisotropic and often categorized as two-dimensional solids. The diversity of observed physical properties originate, in part, from the existence of non-bonding *d* bands and the degree to which they are filled. These materials received significant attention in 1970's and 1980's, being the first family of compounds to exhibit quasi-2D charge-density wave (CDW) transitions.[2,9] In some cases this transition has been related the Fermi surface instability and electron-phonon interaction, similar to Peierls instability in one-dimension. However, in some cases, the transitions are also strongly influenced by electron-electron interaction, like in the Mott phase in 1T-TaS$_2$.[2,9-13] Band-Jahn-Teller and excitonic instabilities have been also suggested in TMDs.[14]

The possibility to intercalate various atoms and molecules[1,15,16] between the TMD layers increases the variety of systems and the diversity of physical properties that may be achieved. The intercalation is generally accompanied by charge transfer between the intercalant species and host layers, providing a tool to fine-tune the electron occupation of the relatively narrow $d$ bands. In the case of Co$_{1/3}$NbS$_2$, cobalt atoms occupy octahedral positions between the triangular prismatic layers of the parent compound 2H-NbS$_2$, resulting in a $\sqrt{3} \times \sqrt{3}$ superlattice.[7] It is thought that two electrons per Co atom are transferred to the initially half–filled d$_{z^2}$ – band of the host NbS$_2$ which thereby becomes 5/6 filled.[2] However, the picture may be more complicated, as witnessed in close relative Co$_{1/3}$NbSe, where several bands contribute to the Fermi surface.[17]

The remaining seven electrons on the Co ion are localized and form magnetic moments with quenched orbital contribution, implying 3/2 spin-only moment.[7,18] These magnetic moments order in an "antriferomagnetic" (AFM) fashion below the temperature $T_N$ of 26 K, producing the hexagonal order of the first kind.[7] The signature of this magnetic order is clearly observed by measurement of the magnetic susceptibility[1] and neutron diffraction experiments.[7] The orthohexagonal magnetic supercell is twice as large as the crystallographic unit cell containing two Co atoms. It may be worth recalling here that antiferromagnetic coupling on a hexagonal lattice usually results in nontrivial magnetic orderings. Indeed, the type of magnetic orderings in other compounds $M_{1/3}NbS2$ (M=Fe, Mn, Ni), based on the same parent compound are known to vary. Different, often competing, magnetic couplings have been considered between intercalated ions. The competition between RKKY and the superexchange via neighboring sulphur ions have been argued in $Co_{1/3}NbS_2$.[4]

It is worth recalling that the parent compound 2H-$NbS_2$ is metallic above and superconducting below $T_c \approx 6$ K. Distinct from its sister compounds (2H-$NbSe_2$, $TaS_2$ and $TaSe_2$) 2H-$NbS_2$ does not undergo a charge density wave transition.[19] Electronic band structure calculations suggests that Fermi level is situated at the middle of the ~ 1 eV wide $d_{z2}$ – band, split from the rest of the d bands by a gap of ~ 1 eV. $T_c$ exhibits no significant change up to pressures of 10 kbar which agrees well with the observed stiffness of the *ab* plane.[20] In contrast, the CDW compound 2H-$NbSe_2$ shows smaller stiffness in the *ab* plane and consequently both transitions are pressure dependent. By increasing the pressure the $T_c$ (7,1 K at 1 bar) increases while the $T_{CDW}$ (35 K at 1 bar) decreases. Such behavior is found for pressures up to 40 kbar.[21] In both compounds, the coupling between *ab* plane layers in the *c* direction is weak, and the main effect of pressure is on the separation between layers.

## 3. EXPERIMENTAL

Single crystals of $Co_{1/3}NbS_2$ were grown from the synthesized charge by means of iodine vapor transport adapting a previously established procedure.[1] Special care was taken to obtain large single crystals which are ideally suited for preset measurements. Samples were cut in a rectangular form of a typical size $0.03 \times 1 \times 3$ mm$^3$. As expected, larger crystals are of a somewhat lower quality. This is evident from the residual resistivity ratio (RRR, the ratio between the resistivity at 300 K and the extrapolated value for $\rho$ at 0 K) of 1.7, which is a slightly lower then RRR of 3 reported for the highest quality samples in Ref. 1. Electrical contacts were painted and golden wires attached to the electrical pads by DuPont silver paste 4299-N. The sample was mounted on a home-made sample holder, which fits into the self-clamped pressure cell. The temperature gradient was generated by heaters at the end of the holder, and measured by a chromel-constantan differential thermocouple. Pressure was monitored *in situ* by a calibrated InSb pressure gauge. The pressure medium used was kerosene. The utility and high precision of this high-pressure experimental setup was already confirmed by measuring other materials.[22,23] Thermal conductivity was measured using an absolute steady-state heat-flow method. The thermal flux through the sample was generated by a 1 kΩ RuO$_2$ chip resistor, glued to one end of the sample, while the other end was attached to the copper heat sink. The temperature gradient across the sample was measured by a chromel-gold with 0.07 at.% Fe differential thermocouple.

## 4. RESULTS AND DISCUSSION

### 4.1. Ambient pressure

The dc-resistivity $\rho(T)$ exhibits a monotonous metallic behavior in the whole measured temperature range, as shown in Fig. 1. The value of the dc-resistivity, $\rho(300\ \text{K}) = 2.4$ μΩm, is larger, but not substantially, from the values found in other transition metal dichalcogenides that are metallic at room temperature. The kink at 26 K corresponds to the magnetic transition of the localized Co spins.[1,24] The presence of such a kink in resistivity indicates that the coupling between the charge carriers in $NbS_2$ planes and localized magnetic moments on Co atoms is not negligible. Upon decreasing the temperature below $T_N$ the slope of $\rho(T)$, becomes steeper. This has been attributed to the decrease of a scattering on spin disorder, pronounced above $T_N$ and reduced upon spin-ordering.[24]

There are just a few measurements of the thermal conductivity $\kappa$ in the TMD's.[25] The very first for $Co_{1/3}NbS_2$ is shown in Fig. 2a. Already at the first sight site it may be noted that, otherwise surprisingly featureless, $\kappa(T)$ exhibits two different temperature regimes. The separation of electronic ($\kappa_{el}$) and phonon ($\kappa_{ph}$) contributions is done according the Wiedemann-Franz (WF) law, $\kappa_{el}(T) = \left(\pi^2 k_B^2 T / 3e^2\right)\sigma(T)$. Here $\sigma(T)$ is the electrical conductivity related to the resistivity $\rho(T)$, presented in Fig. 1, via $\sigma(T)=1/\rho(T)$. It is important to recall that the WF-law[26] is valid if the elastic scattering of electrons is the dominant scattering process. This is usually realized at temperatures above the Debye temperature ($\theta_D$).[27] In case of $NbS_2$ $\theta_D \sim 260$ K.[28-30] At lower temperatures, $T<\theta_D$, WF-law is applicable in materials where the residual resistivity, originating from the elastic scattering on defects, is high. Since our $Co_{1/3}NbS_2$ samples show large residual resistivity of 136 μΩcm and a small positive temperature coefficient, the validity of the WF law can be assumed. $\kappa_{el}$ is about ~3 W/mK at room temperature (RT). At the same temperature, the ratio $\kappa_{el}/\kappa_{tot}$ is approximately 0.36, implying that electrons carry about one-third of the heat. The weak linear increase of thermal conductivity above 100 K, $d\kappa/dT \approx 0.01$ W/mK$^2$, is essentially due to electrons, while phonon contribution is almost temperature independent. Upon lowering the temperature the small $\kappa_{el}/\kappa_{tot}$ ratio becomes even smaller, showing a minimum around the magnetic transition temperature $T_N$. Consequently, the anomaly in $\kappa_{el}$, originating from the one in $\rho(T)$ around $T_N$, turns out to be very small and practically invisible in Fig. 2a. In the whole temperature range the dominant contribution to the thermal conductivity is, as shown in Fig. 2a, the phonon contribution $\kappa_{ph}$. The low temperature (T<10 K) rise of $\kappa_{ph}(T)$ follows the power low ($T^\nu$) behavior with $\nu \approx 1.4$. This usually speaks in favor of significant disorder in the system,[31] but probably should not be considered independently from the behavior that follows at higher temperature. As seen from the graph, the rise of $\kappa_{ph}(T)$ is followed by a smooth saturation at 5 W/mK for temperatures above above 130 K. The lack of the classical Umklapp maximum in $\kappa_{ph}$ is probably related to this rather low value of $\kappa_{ph}$ at high temperature, demonstrating a significant scattering of acoustic phonons. Although the source of the scattering that leads to the saturation and a low value of $\kappa_{ph}(T)$ remains unknown, disorder is considered to be the most likely cause.[32]

Yet another possibility in the discussion of the thermal conductivity is that the scattering on Co spins may play an important role, similar to that observed in charge transport $\rho(T)$. Such a claim is not supported by our data, since $\kappa_{tot}(T)$ does not show a significant rise or an overall change of regime below $T_N$. However, weak effects of magnetic ordering can be identified in the derivative of the thermal conductivity data, shown in Fig. 2.b). Interestingly, the effect of magnetic ordering is much more pronounced in $d\kappa_{tot}(T)/dT$ than in $d\kappa_{el}(T)/dT$. The qualitative difference in the two derivatives indicates that the feature seen in the derivative of the total thermal conductivity at magnetic transition temperature is not a pure electronic effect, but can be attributed to the lattice and/or magnetic degrees of freedom. The absence of clear rise in $\kappa_{ph}$ upon ordering at 26 K also counters the possibility that a *structural* ordering of cobalt ions takes place at the critical temperature. This type of ordering has not been reported for cobalt system, though it is argued to occur, for example, in silver intercalated system $Ag_xNbS_2$, for example, where the downturn in resistivity upon cooling was observed at a much higher temperature.[33,34]

Temperature behavior of the thermoelectric power $S(T)$ at ambient pressure is shown in Fig 1. Several features should be noted: Firstly, the value of $S(T)$ at room temperature is rather high for a metal, -20 µV/K., and further increases upon cooling, reaching a maximum of -30 µV/K around 80 K. Since the resistivity is metal-like in the whole temperature range it is tempting to decompose $S(T)$, above 80K, in the following way: $A+BT$. Such a behavior is characteristic for metals, though the intercept A= -36 µV/K is unusually high. Secondly, the negative value of thermoelectric power has an opposite sign to that of the Hall coefficient, which is positive for the whole temperature range.[1] For low temperatures, it should be observed that the inflection point in $S(T)$ (or, the maximum in $dS/dT$) coincides with the magnetic ordering temperature as observed in resistivity. The change of sign in $S(T)$ appears inside the ordered phase. A more detailed discussion of $S(T)$ in the low temperature region is left for the next section, since this is also where the effect of pressure is most pronounced.

The features in $S(T)$ listed above suggests a rather complicated Fermi surface in $Co_{1/3}NbS_2$, which accommodates the portions with electron- and hole-like electron dispersions, i.e. exhibiting quite different effective masses. The direct measurements and detailed electronic band structure calculations that would detail this are missing at present. However, the existing measurements on sister compounds $M_{1/3}NbS_2$ and the calculations for $Co_{1/3}NbS_2$ suggest that the band in $NbS_2$ layer stays stiff upon intercalation. It also suggests that the Fermi level is determined only by the amount of the charge transfer from intercalated ions to $NbS_2$ layer, and consists of several parts with different dispersions.[17]

For high temperatures, the reason for the maximum at 80 K cannot be explained by the phonon drag which is as the usual suspect since the temperature of the maximum is much higher than expected $\theta/5$. Moreover, the low overall value $\kappa_{ph}(T)$ suggests a rather weak thermal flow of acoustic phonons, while the usual *Umklapp*-related decrease of $\kappa_{ph}(T)$ at high temperatures, qualitatively related to phonon-drag-maximum in $S(T)$, is lacking. On the other hand, it may be remarked that the unusual high zero temperature intercepts of the extrapolated high temperature linear behaviors are also observed in other metallic-like materials, e.g. high-$T_C$ superconductors[35] or quasi-one dimensional systems (e.g. $BaVS_3$).[36] The parent material $2H$-$NbS_2$ also develops quasi-constant term in $S(T)$ above 100 K, although its value is considerably smaller than in $Co_{1/3}NbS_2$.[34]

4.2. High pressure

The study of the $Co_{1/3}NbS_2$ under pressure reveals several interesting features. Figs. 3 and 5 show how dc-resistivity and thermoelectric power change under pressure, below 40 K. The behavior of dc-resistivity under pressure is rather complicated, indicating that the behavior of the system is determined by a subtle interplay of competing interactions. As the pressure increases, the room temperature value of the resistivity decreases while the residual resistivity rises and reaches a maximum at 11.2 kbar, followed by a decrease at even higher pressures. Simultaneously, the kink in $\rho(T)$, previously identified as marking the magnetic ordering, shifts to lower temperature, indicating the expansion of the high temperature phase on the account of the low temperature phase. Since the strength of the kink weakens with increasing pressure, the easiest way to determine the precise position of the kink is through the derivative $d\rho(T)/dT$. This is shown in Fig. 4 where the beginning of the upturn in $d\rho(T)/dT$ reveals the transition temperature $T_N$.

As shown in the previous section, the thermoelectric power $S(T)$ at ambient pressure has several prominent features below 40 K (see Fig. 1). Instead of monotonously falling towards zero upon lowering the temperature, $S(T)$ first changes sign (becomes positive) around 18 K, then exhibits a hump around 11 K, and finally approaches zero from the positive side. It is important to note that the maximum at 11 K has nothing to do with the phonon-drag effect. Here, the temperature of the maximum is much lower the usual $\theta_D/5$ (estimated to be 50 K from the Debye temperature in the parent compound 2$H$-NbS$_2$,[29]). Another way to confirm that the observed hump in $S(T)$ is not related to the phonon drag is to apply pressure. As shown in the Fig. 5 the hump is strongly pressure dependent and fully vanishes as the pressure is increased. This rules out the phonon drag explanation which is expected to be weakly dependent on pressure.

We emphasize again that the $S(T)$ dependence is steepest exactly at the $T_N$ (indicated by the arrow on Fig. 1.) The method to determine $T_N$ from the maximum of the derivative $dS/dT$ (see Fig. 6) is interchangeable with the one using position of the kink in $\rho(T)$, i.e. two methods result in the same pressure dependence of $T_N$. The observed decrease of the ordering temperature with pressure decreases at the rate $dT_N/dP \sim -1$ K/kbar. The resulting T-P phase diagram is shown as inset in Fig. 5. We conclude that anomalous low temperature behavior in $\rho(T)$ and $S(T)$ is, with no doubt, a direct consequence of the magnetic ordering in $Co_{1/3}NbS_2$, and the related interplay between the magnetically active localized Co spins and the itinerant electrons.

## 5. DISCUSSION

Ref. 4 argues that two competing mechanisms may result in the magnetic ordering at $T_N$: namely the RKKY interaction and the super-exchange interaction via neighboring sulphur ions. With pressure, which brings the layers closer, the overlap between cobalt and sulphur orbitals is expected to increase, thus increasing the superexchange coupling, as well the coupling of the cobalt spin state to the conducting electrons in the layer and the RKKY interaction.

The effect of pressure on $Co_{1/3}NbS_2$ is particularly puzzling in two respects: First, the *increase* of resistivity with increasing pressure in the metallic region that extends in a wide temperature range above and below room temperature is an effect unseen in TMD's. However, it may be envisaged that the increase of the coupling, and thus of the scattering, of conducting electrons on disordered spins which reside on cobalt atoms, may be the responsible mechanism. This goes hand in hand with the suggestion by Friend *et al.*[1] that this type of scattering is pronounced at temperatures above the ordering temperature. It may be noted however, that effect has not been observed in the sister compound, $M_{0.05}TiS_2$ (M=Co, Fe, Ni) where the usual decrease of resistivity with pressure was observed.[37] The reason may lie in the much lower concentration of magnetic ions in the latter case. The lower concentration implies lower scattering rate on spin-disorder at high temperature, as well as the absence of the magnetic superstructure at low temperature.

The increase of the magnetic coupling that probably causes the negative sign of $d\rho/dp$ at high temperature in $Co_{1/3}NbS_2$ is at odds with the second observed effect of pressure, namely the decrease of $T_N$, the magnetic ordering temperature, with increasing pressure. Several reasons come to mind as being responsible for this behavior. Competing super-exchange (antiferromagnetic) and RKKY (ferromagnetic) interactions increase at different rates under pressure. The pressure-induced increase of the coupling of the Co-spin to conducting electrons, as suggested by the high temperature data, then primarily implies an increase in the RKKY interaction. The assumption that the superexchange interaction is less affected implies a diminishing $T_N$ with pressure.

However, the above picture is not the only one which can explain the observed phenomena. A possibility is that the spin on Co site gets screened by conducting electrons as the coupling increases, and the temperature lowers. This goes along the lines of the mechanism for reducing the magnetic ordering under pressure in Kondo-lattice systems, proposed long ago by Doniach.[38] As such, behavior of $T_N$ with pressure should follow the classical Doniach's phase diagram.[38,39] Indeed, the observed decrease of $T_N$ resembles quite well the part of Doniach's diagram where the Kondo screening prevails over the, in this case, antiferromagnetic RKKY interaction, i.e. where $T_K > T_{RKKY}$. It should be noted that whether the RKKY interaction is antiferromagnetic or ferromagnetic depends on the filling of several bands that cross the Fermi level, which, in case of $Co_{1/3}NbS_2$, is not unambiguously determined.

The change of the spin state of Co atom caused by the variation of the separation between layers has also been suggested by recent experiments by Nakayama et al.[18] In their experiments, the concentration of the Co atoms is varied, $Co_xNbS_2$ ($0.15 \leq x \leq 0.55$). The monotonous decrease of the lattice constants with decreasing $x$ is accompanied by a drop of the Co-spin form 3/2 to 1/2 as $x$ falls below 0.33. Apart from the Kondo screening, mentioned above, the change of the spin state, especially at high temperature, may be related to the change of the Madelung crystal field at the Co site. Consequently, a change of the Co charge may not be ruled out, since electronic states of cobalt appear close to the Fermi level in band-structure calculations reported in the same paper. These calculations indicate a sizable hybridization of Co-states and conducting band, represented by the share of the Co-states in the total density of states (DOS) and the width of the Co-projected DOS.

Whatever the mechanism behind the change of the spin state in $Co_xNbS_2$, it seems probable that doping below $x \approx 0.33$, and the application pressure in the stoichiometric compound may cause the same instability in the system.

## 6. CONCLUSION

Our measurements of the temperature and pressure dependence of the transport properties of $Co_{1/3}NbS_2$ reveal unexpected effects, particularly regarding the behavior of the pressurized system near the magnetic ordering temperature. The explanation of the observed effects is primarily due to the pressure-induced change of the coupling of Co-spin state to conducting electrons. Our results indicate that the pressure in $x = 1/3$ compound touches the delicate balance between magnetic interaction and spin states motivating further experimental work, to probe more directly the spin and charge states in pressurized system. Further theoretical work is required to determine the precise band structure which could reveal the role of the possible RKKY interaction. In this sense the measurements presented in this paper are of great importance since we have established the *p-T* phase diagram based on two independent physical properties, thermoelectric power and dc-resistivity. Furthermore, from the measurements of the thermal conductivity it follows that transition at 26 K is not related to structural ordering of Co. However, the analysis of $d\kappa_{tot}(T)/dT$ and $d\kappa_{el}(T)/dT$ reveals an interplay of lattice and magnetic degrees of freedom. Rather complicated pressure dependences of dc-resistivity and thermoelectric power indicate the presence of competing interactions in the system, while the opposite signs of S and the Hall Effect point to a subtle topology of the Fermi surface consisting of several parts with different dispersions.

**Acknowledgements**

Valuable discussions with Ivo Batistić are acknowledged. This work was supported by the Unity through Knowledge Fund, under grant No. 65/10, and Croatian Ministry of Science, Education and Sports grants 035-0352826-2848, 035-0352826-2847, and 177-0352826-047. NB acknowledges the National Foundation for Science, Higher Education and Technological Development of the Republic of Croatia and the Alexander von Humboldt foundation.

**Figure captions:**

**Fig. 1.** (Color online) The ambient pressure dc-resistivity, $\rho(T)$ (○), and thermoelectric power, $S(T)$ (◇), of $Co_{1/3}NbS_2$. Temperature $T_N$, at which antiferromagnetic ordering sets in, is marked by an arrow.

**Fig. 2.** (Color online) a) Measured thermal conductivity, $\kappa_{tot}(T)$ (○), is decomposed in lattice, $\kappa_{ph}(T)$ (△) and electronic contributions $\kappa_{el}(T)$ (--). b) Temperature dependence of $d\kappa_{tot}(T)/dT$ (○) and $d\kappa_{el}(T)/dT$ (▽) in the narrow temperature range around $T_N$ reveals an involvement of structural degrees of freedom in the transition.

**Fig. 3.** (Color online) a): Pressure dependence of the low temperature dc-resistivity. b) dc-resistivity in the whole measured temperature range for selected values of pressure. Note that the resistivity increases with increasing pressure in the high temperature region.

**Fig. 4.** (Color online) Pressure dependence of the derivative of electric resistivity, $d\rho/dT$ in the low temperature region. The temperature of magnetic ordering corresponds to the minimum of the derivative.

**Fig. 5.** (Color online) *Main panel*: The temperature dependence of the thermoelectric power and its pressure behavior in the low temperature range. *Inset*: The pressure dependence of the transition temperature $T_N$, as determined from the minimum of the derivative $d\rho/dT$.

**Fig. 6.** (Color online) Pressure dependence of the derivative of the thermoelectric power, $dS/dT$, in the low temperature range. $T_N$ is identified with the minimum of $dS/dT$ (i.e., the temperature for which the temperature dependence of $S(T)$ is the steepest).

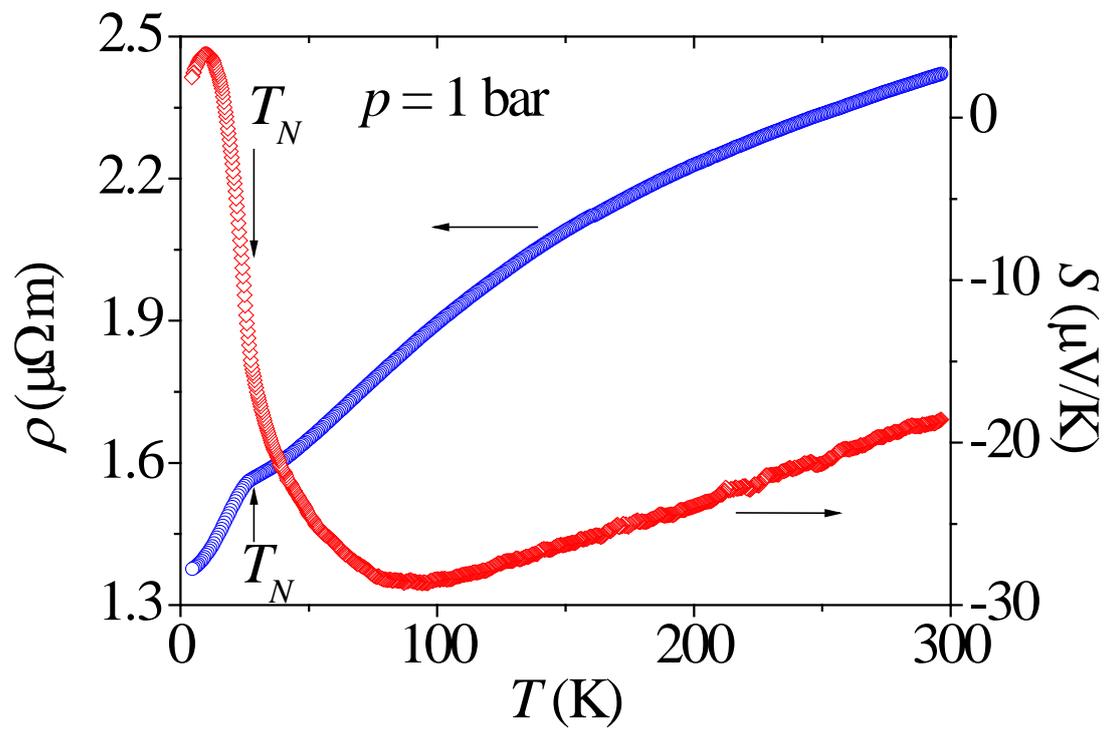

Fig. 1

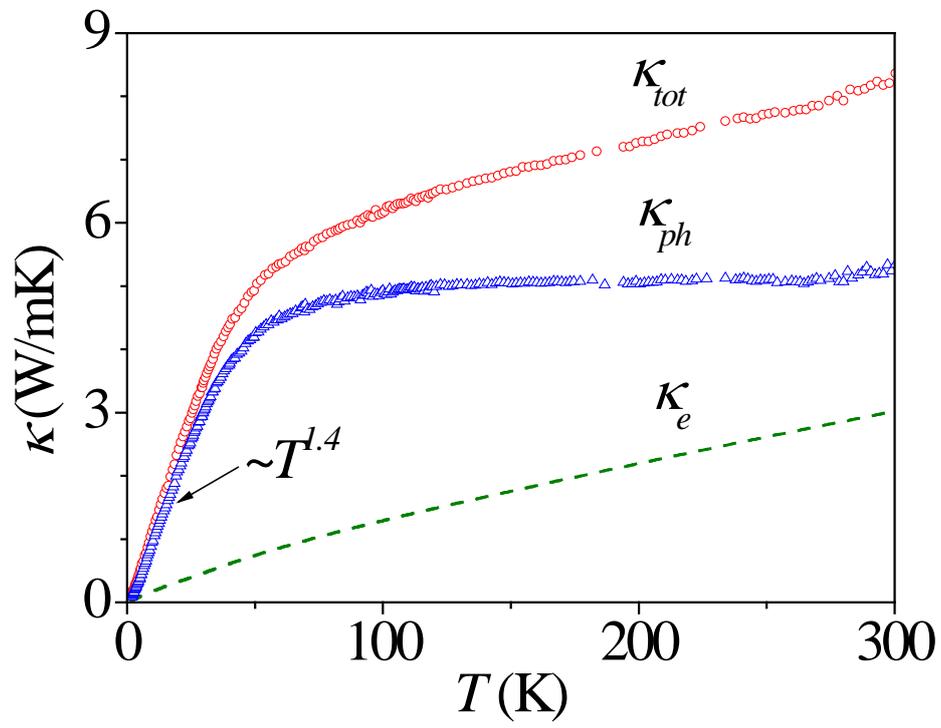

Fig. 2a

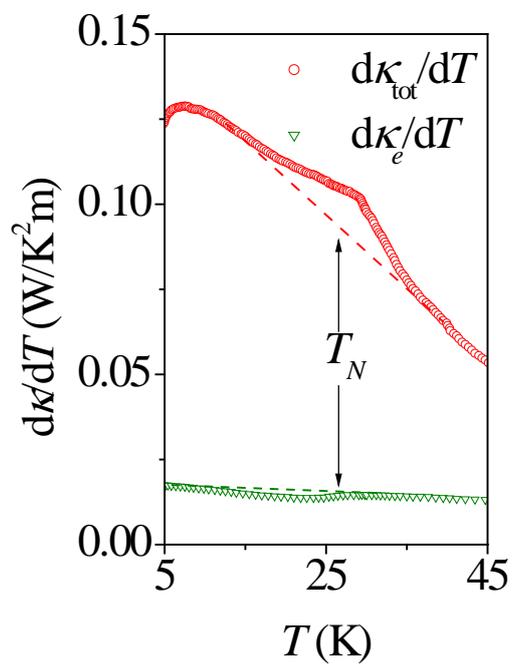

Fig. 2b

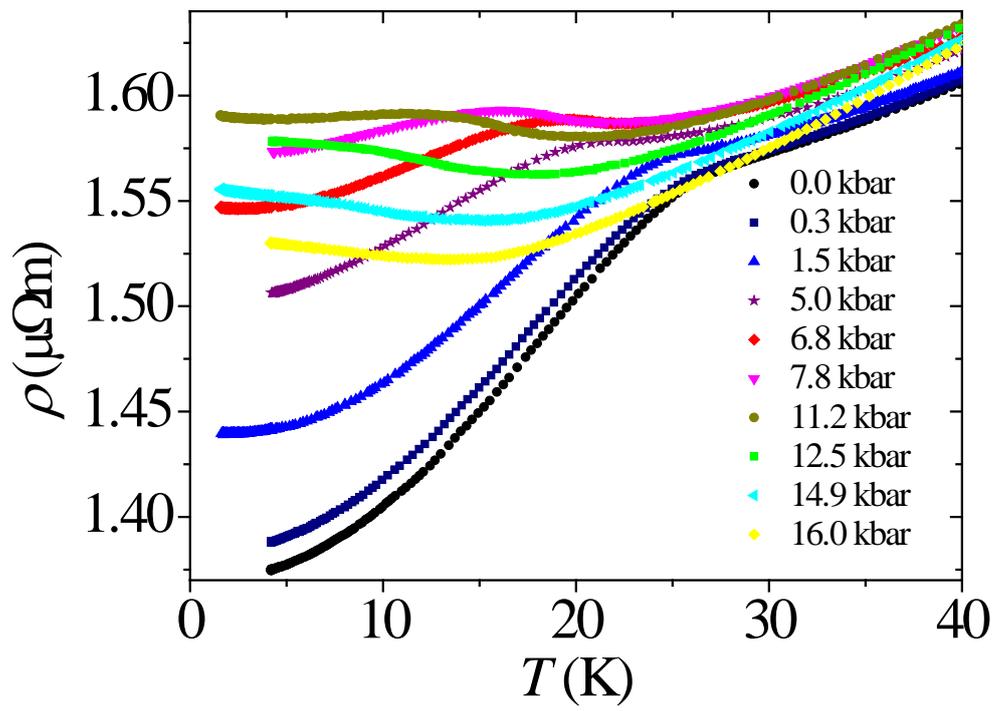

Fig. 3a

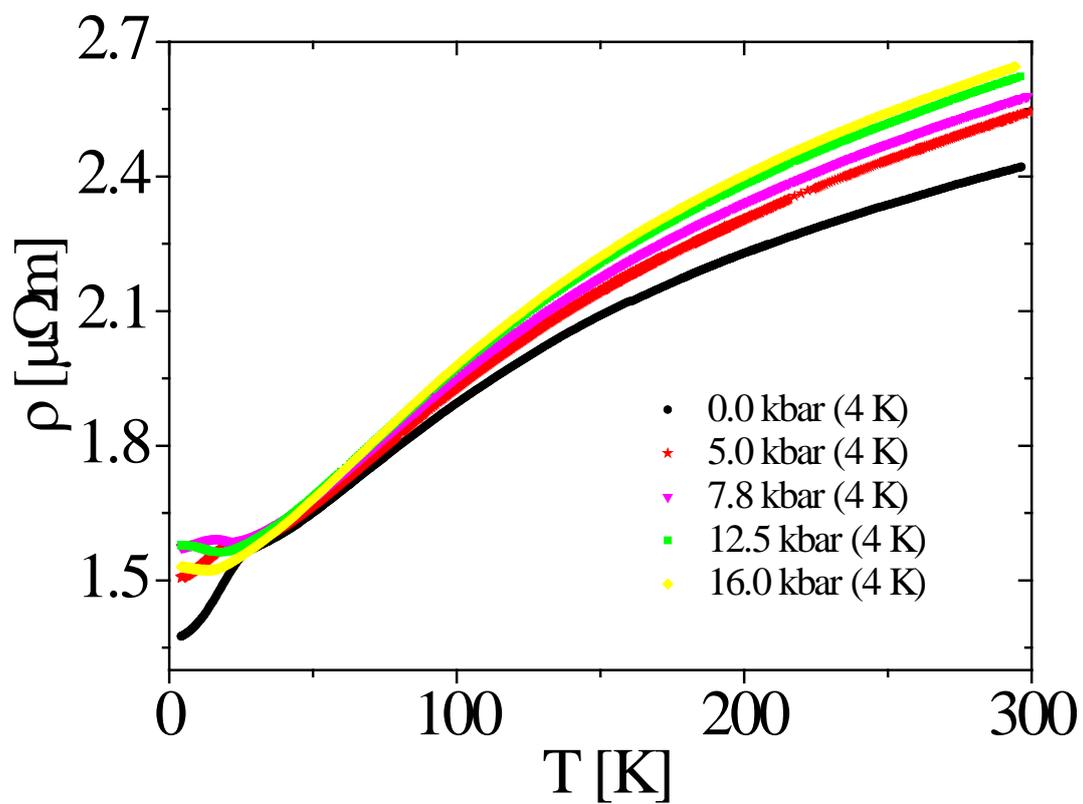

Fig. 3b

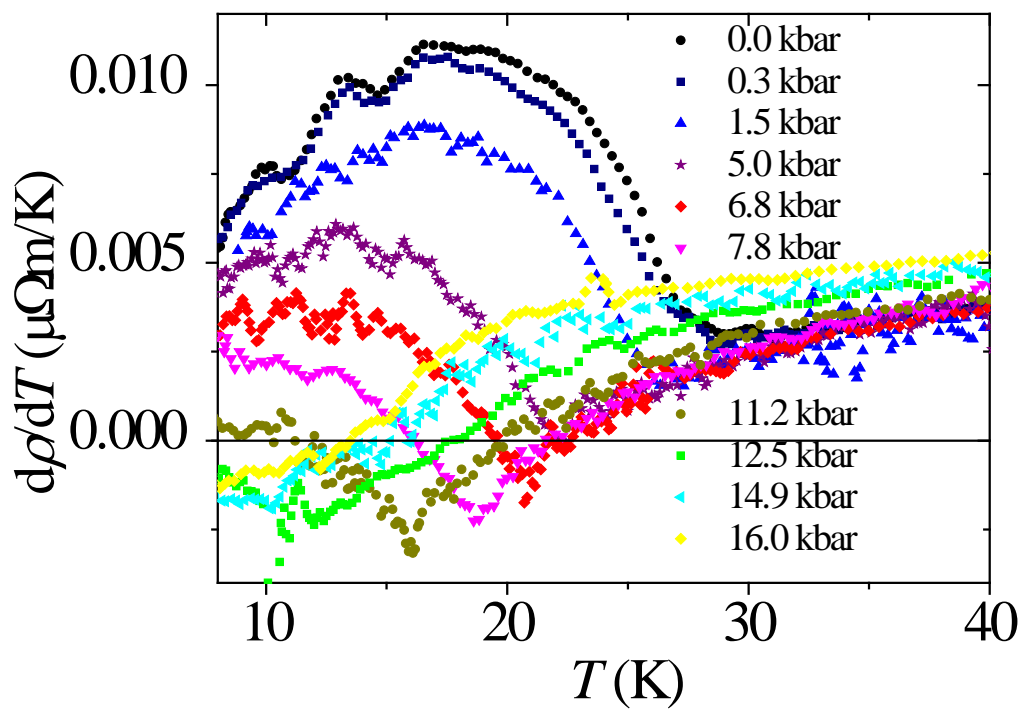

Fig. 4

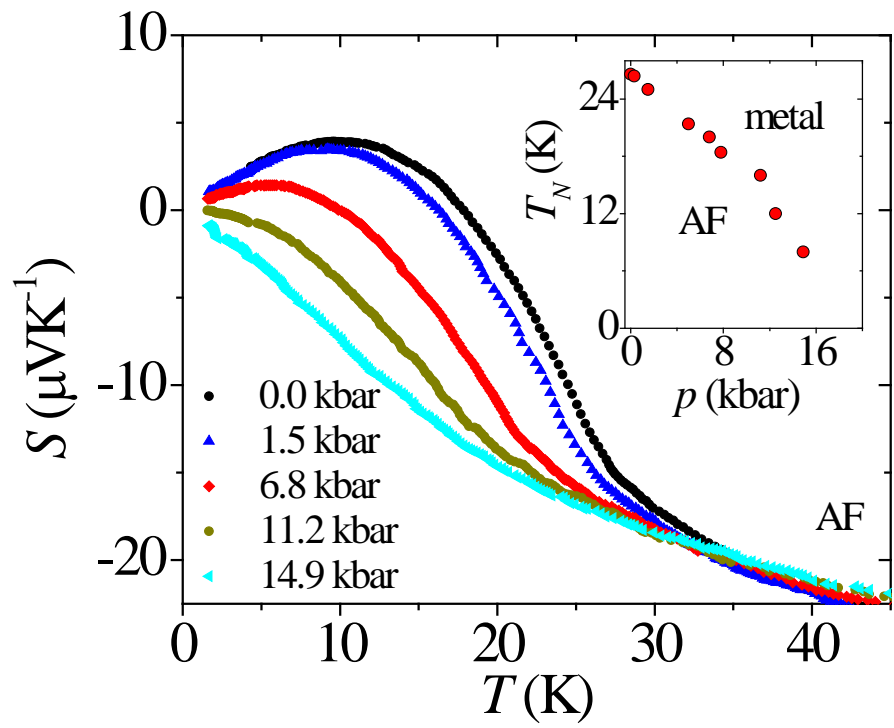

Fig. 5

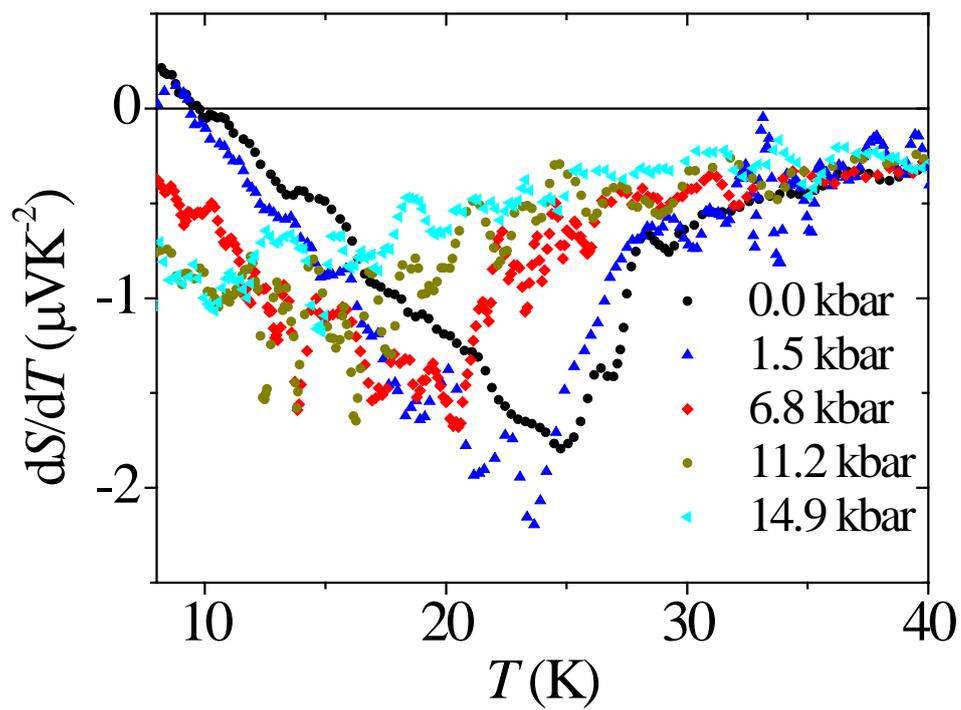

Fig. 6